\documentclass[prd,showpacs,showkeys,nofootinbib,floatfix,eqsecnum,
               fleqn,preprint,12pt,tightenlines]{revtex4-1} %for preprint

\usepackage{amsmath,amssymb,revsymb,graphicx,dcolumn}

\newcommand{\beq}{\begin{equation}}
\newcommand{\eeq}{\end{equation}}
\newcommand{\beqa}{\begin{eqnarray}}
\newcommand{\eeqa}{\end{eqnarray}}
\newcommand{\bsubeqs}{\begin{subequations}}
\newcommand{\esubeqs}{\end{subequations}}

\newcommand\ddfracNEW[2]{\displaystyle{\frac{#1}{#2}}}             %%better

\begin{document}

\begin{widetext}
\noindent Phys. Rev. D {\bf 100}, 083534 (2019) \hfill  arXiv:1904.09961
%
%\noindent arXiv:1904.09961 \hfill KA--TP--06--2019\;(\version)
%
%\noindent  \hfill KA--TP--06--2019\;(\version)
\newline\vspace*{5mm}
\end{widetext}

\title{Nonsingular bouncing cosmology from general relativity}%
\vspace*{1mm}

\author{F.R. Klinkhamer}
\email{frans.klinkhamer@kit.edu}

\affiliation{Institute for Theoretical Physics,
Karlsruhe Institute of Technology (KIT),\\
76128 Karlsruhe, Germany\\}

\author{Z.L. Wang}
\email{ziliang.wang@kit.edu}

\affiliation{Institute for Theoretical Physics,
Karlsruhe Institute of Technology (KIT),\\
76128 Karlsruhe, Germany\\}

\begin{abstract}
\vspace*{1mm}\noindent
We investigate a particular type of classical nonsingular bouncing cosmology,
which results from general relativity if we allow for degenerate metrics.
The simplest model has a matter content with a constant
equation-of-state parameter
and we get the modified Hubble diagrams for both
the luminosity distance and the angular diameter distance.
Based on these results, we present a \textit{Gedankenexperiment}
to determine the length scale of the spacetime defect which has replaced
the big bang singularity.
A possibly more realistic model has an equation-of-state parameter
which is different before and after the bounce. This last model also
provides an upper bound on the defect length scale.
\vspace*{-0mm}
\end{abstract}


\maketitle

\section{Introduction}
\label{sec:Intro}

There have been discussions from various physics perspectives
of the possible existence of a pre-big-bang phase, with or without
a bounce-type behavior of the cosmic scale factor. See, e.g.,
Refs.~\cite{GasperiniVeneziano2002,AshtekarSingh2011,IjjasSteinhardt2018}
and references therein.

Recently, we have obtained
a surprising hint for the actual existence
of a pre-big-bang phase~\cite{Klinkhamer2019},
where we worked with the established theory of general relativity in
four spacetime dimensions but allowed for degenerate metrics.
(See the last two paragraphs of Sec.~I in
Ref.~\cite{Klinkhamer2019} for a brief comparison of this extended
version of general relativity and the standard version,
which considers only nondegenerate metrics.)
With an appropriate differential structure and trivial
spacetime topology, a nonsingular spatially flat
Friedmann-type solution of the Einstein gravitational field equation
has been obtained, where the curvature
and the matter energy density remain finite
(these quantities diverge for the standard Friedmann solution).
Most interestingly, this nonsingular
Friedmann-type solution suggests the existence
of a ``pre-big-bang'' phase (in standard terminology)
with a bounce-type behavior of the cosmic scale factor.

The aim of the present article is to review this nonsingular bounce,
which remains within the realm of general relativity,
and to obtain a better understanding of the
nonsingular bouncing cosmology by performing
exploratory calculations of certain cosmological observables.
Even though, at this moment, these cosmological observables
are only accessible through \textit{Gedankenexperiments},
it is instructive to discuss them. In the Appendix,
we also give an explicit realization
of a particular classical nonsingular bouncing cosmology that was
discussed in Ref.~\cite{IjjasSteinhardt2018}
(this cosmology has a different matter content before and
after the bounce). The model of the Appendix
allows us to obtain a qualitative upper bound on
the length scale of the spacetime defect which has replaced
the big bang singularity.

%%\newpage%%tmp
\section{Nonsingular bounce with a constant equation of state}
\label{sec:Nonsingular bounce-with-a-constant-EOS}

We start from the classical spacetime manifold of Ref.~\cite{Klinkhamer2019},
but use a simplified version of the cosmic time coordinate $T$ and
consider only the $T$-even solution for the cosmic scale factor $a(T)$.
In this way, we obtain a modified spatially flat
Friedmann--Lema\^{i}tre--Robertson--Walker (FLRW)
universe with a bounce-type behavior of $a(T)$.
We can be relatively brief in this section, as further details can be found
in Refs.~\cite{Klinkhamer2019,Klinkhamer2014-mpla,KlinkhamerSorba2014,%
Guenther2017}.
Throughout, we use natural units with $c=1$ and $\hbar=1$.

With a cosmic time coordinate $T$ and comoving
spatial Cartesian coordinates $\{x^{1},\,  x^{2},\, x^{3}\}$,
an appropriate \textit{Ansatz}
for the metric is given by~\cite{Klinkhamer2019}%
\bsubeqs\label{eq:mod-FLRW}
\beqa\label{eq:mod-FLRW-ds2}
\hspace*{-6mm}
ds^{2}\,\Big|_\text{mod.\;FLRW}
&\equiv&
g_{\mu\nu}(x)\, dx^\mu\,dx^\nu \,\Big|_\text{mod.\;FLRW}
=
- \frac{T^{2}}{b^{2}+T^{2}}\,dT^{2}
+ a^{2}(T)
\;\delta_{kl}\,dx^k\,dx^l\,,
\\[2mm]
\hspace*{-6mm}
b &>& 0\,,
%\\[2mm]
\eeqa\beqa
\hspace*{-6mm}
\phantom{s^{2}\,\Big|_\text{mod.\;FLRW}}
T &\in& (-\infty,\,\infty)\,,
%\\[2mm]
\eeqa\beqa
\hspace*{-6mm}
\phantom{s^{2}\,\Big|_\text{mod.\;FLRW}}
x^k &\in& (-\infty,\,\infty)\,,
\eeqa
\esubeqs
where the function $a(T)$ in  \eqref{eq:mod-FLRW-ds2}
is, strictly speaking, not the same as the function $a(\tau)$
in (3.6) of Ref~\cite{Klinkhamer2019}.
The parameter $b$ in the metric \eqref{eq:mod-FLRW-ds2}
corresponds to the characteristic length scale
of the spacetime defect localized at $T=0$
(see Refs.~\cite{Klinkhamer2019,Klinkhamer2014-mpla,%
KlinkhamerSorba2014,Guenther2017} and references therein).
For the moment, $b$ is simply a model parameter
and we remain agnostic as to its physical origin.
It may be that $b$ is related to the Planck
length, but it is also possible that $b$
is related to a new fundamental length scale
of quantum spacetime~\cite{Klinkhamer2007}.

We observe that the metric \eqref{eq:mod-FLRW-ds2} is degenerate,
with $\det\,g_{\mu\nu} = 0$ at $T=0$.
The corresponding $T=0$ spacetime slice
may be interpreted as a 3-dimensional ``defect'' of spacetime
with topology $\mathbb{R}^3$.
The standard elementary-flatness condition does not hold
at the location of this spacetime defect;
see App.~D in Ref.~\cite{Klinkhamer2014-mpla}
and Sec.~2~D  in Ref.~\cite{KlinkhamerSorba2014} for further discussion.
As will be seen shortly, the metric \eqref{eq:mod-FLRW-ds2}
removes the big bang curvature singularity, but does so at the
price of introducing a spacetime defect.
We also remark that a degenerate metric evades certain singularity theorems;
cf. Sec.~3.1.5 in Ref.~\cite{Guenther2017}.

\begin{figure}[t]
\vspace*{0mm}
\begin{center}
\includegraphics[width=0.5\textwidth]{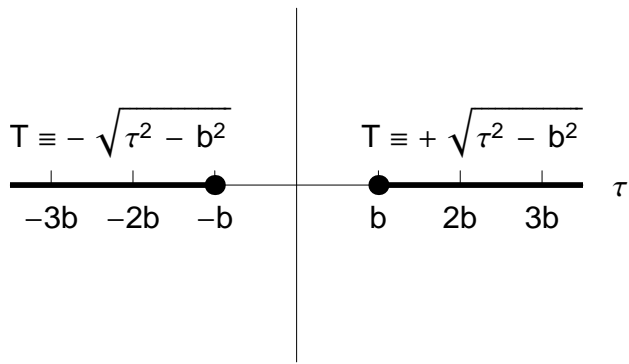}
%%{messengers-from-a-pre-BB-phase-fig01-v022.eps}
\end{center}
\vspace*{-5mm}
\caption{Surgery on the real line with coordinate $\tau \in \mathbb{R}$
gives the cosmic time axis
$\tau \in  (-\infty,\,-b]  \, \cup \, [b,\,\infty)$,
where the points $\tau=-b$ and $\tau=b$
are identified (as indicated by the dots).
A suitable cosmic time coordinate is given by $T \in \mathbb{R}$
from \eqref{eq:T-def-pos-neg}.
Each point of the cosmic time axis corresponds to
a unique value of the coordinate $T$.
}
\label{fig:sketch-tau-axis-with-T}
\vspace*{0mm}
\end{figure}

Later on, we will simplify the calculations away from the spacetime defect
by use of the auxiliary coordinate $\tau$ instead of $T$.
These two coordinates are related
as follows (Fig.~\ref{fig:sketch-tau-axis-with-T}):%
\bsubeqs\label{eq:T-def}
\beqa\label{eq:T-def-pos-neg}
T(\tau) &=&
\begin{cases}
 + \sqrt{\,\tau^{2} - b^{2}}\,, &  \;\;\text{for}\;\; \tau \geq b\,, \\[1mm]
 - \sqrt{\,\tau^{2} - b^{2}}\,, &  \;\;\text{for}\;\; \tau \leq -b \,,
 \end{cases}
\\[2mm]
\tau &\in&  (-\infty,\,-b]  \, \cup \, [b,\,\infty)\,.
\eeqa
\esubeqs
The inverted relation reads
\beqa
\label{eq:mod-FLRW-tau-of-T-def}
\tau(T)&=&
\begin{cases}
 + \sqrt{b^{2}+T^{2}}\,,    & \;\;\text{for}\;\; T \geq 0\,,
 \\[2mm]
 - \sqrt{b^{2}+T^{2}}\,,    & \;\;\text{for}\;\; T \leq 0\,,
\end{cases}
\eeqa
which is multivalued at $T=0$.
The advantage of using the auxiliary coordinate
$\tau$ is that the metric \eqref{eq:mod-FLRW-ds2}
takes the standard spatially flat FLRW form,
$ds^{2} =  - d\tau^{2} + a^{2}(\tau) \;\delta_{kl}\,dx^k\,dx^l$,
and that the reduced
field equations are nonsingular.
But it is important to realize that the coordinate transformation
from $T$ to $\tau$ is not a diffeomorphism
(an invertible $\text{C}^{\infty}$ function by definition):
the function \eqref{eq:mod-FLRW-tau-of-T-def} is discontinuous
between $T=0^{-}$ and $T=0^{+}$,
as is the (suitably defined) second derivative.
In short, the differential structure of the metric \eqref{eq:mod-FLRW-ds2}
in terms of $T$
is different from the one of the standard spatially flat FLRW metric
in terms of $\tau$;
see Ref.~\cite{KlinkhamerSorba2014} for a related discussion.

%
%% rename: fig0123456-v1.eps=fig0123456-v2.eps
%% rename: bounce-cosmology-fig01234567-v2.eps=bouncing-cosmology-fig01234567-v3.eps
%
%% rename: bounce-cosmology-fig0167-v3.eps=bouncing-cosmology-fig0167-v4.eps
%
%% relabel fig0N-v4.eps --> figN-v4.eps for N=1,2,3
%% relabel fig06-v4.eps --> fig5-v4.eps
%% relabel fig07-v4.eps --> fig7-v4.eps
%
%% relabel figN-v4.eps --> fig0N-v401.eps     for N=1,2,3,4
%% relabel fig6-v4.eps --> fig07-v401.eps
%
%% relabel fig0N-v401.eps --> fig0N-v5.eps     for N=1,...,7
%

Taking the metric \eqref{eq:mod-FLRW-ds2}
with spacetime coordinates $\{T,\, x^{1},\,  x^{2},\, x^{3}\}$
and the energy-momentum tensor of a homogeneous perfect fluid,
the Einstein equation gives the following
modified spatially flat Friedmann
equation and energy-conservation equation,
together with an assumed equation-of-state parameter $W(T)$:
\bsubeqs\label{eq:mod-Friedmann-equations-abc}
\beqa\label{eq:mod-Friedmann-equation-a}
\left(1+ \frac{b^{2}}{T^{2}}\right)
\left( \frac{1}{a(T)}\,\frac{d a(T)}{d T} \right)^{2}
&=& \frac{8\pi G_N}{3}\;\rho(T)\,,
\\[3mm]
\label{eq:mod-Friedmann-equation-b}
\frac{d}{d a} \bigg[ a^3\,\rho(a)\bigg]+ 3\, a^{2}\,P(a)&=&0\,,
\\[3mm]
\label{eq:mod-Friedmann-equation-c}
W(T) \equiv \frac{P(T)}{\rho(T) } = w  &=& 1\,,
\eeqa
\esubeqs
where the last equation corresponds to a particular choice
for the constant equation-of-state parameter $w$.
The actual value $w = 1$ in \eqref{eq:mod-Friedmann-equation-c}
matches the one used in Ref.~\cite{IjjasSteinhardt2018}
and the reason for this choice will be
discussed further in Sec.~\ref{sec:Discussion}.
As mentioned in Ref.~\cite{Klinkhamer2019},
the only new ingredient in \eqref{eq:mod-Friedmann-equations-abc}
is the singular factor $(1+ b^{2}/T^{2})$ on the left-hand side of the
modified Friedmann equation \eqref{eq:mod-Friedmann-equation-a}.

\begin{figure}[t]
\vspace*{-0mm}\begin{center}
\includegraphics[width=0.5\textwidth]{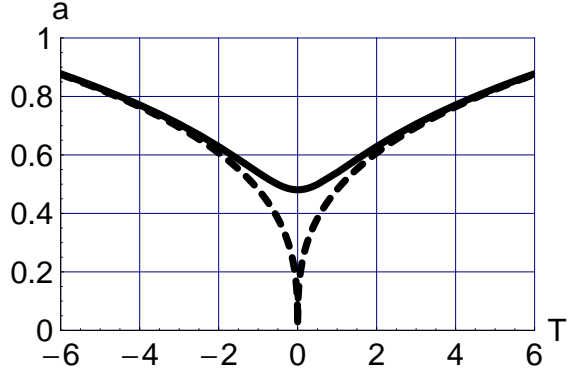}
%%{nonsingular-bouncing-cosmology-fig02-v380.eps}
%%{nonsingular-bouncing-cosmology-fig02-v3.eps}
%%{messengers-from-a-pre-BB-phase-fig02-v022.eps}
\end{center}\vspace*{-5mm}
\caption{Cosmic scale factor (full curve) of the modified spatially flat
FLRW universe with $w=1$ matter,
as given by \eqref{eq:regularized-Friedmann-asol} with $b=1$ and $T_{0}=4\,\sqrt{5}$.
Also shown is the cosmic scale factor (dashed curve) of the standard FLRW universe
with an extended cosmic time coordinate $T$,
as given by \eqref{eq:regularized-Friedmann-asol}
with $b=0$ and $T_{0}=4\,\sqrt{5}$.}
\label{fig:a-bounce-a-singular}\vspace*{0mm}
\end{figure}

The $T$-even bounce-type solution $a(T)$
from \eqref {eq:mod-Friedmann-equations-abc}
with normalization $a(T_{0})=1$ at $T_{0}>0$ is given by
\beq
\label{eq:regularized-Friedmann-asol}
a(T)\,\Big|_\text{mod.\;FLRW}^\text{(w=1,\;$T$-even\;sol.)} =
\sqrt[6]{\big(b^{2}+T^{2}\big)\big/\big(b^{2}+T_{0}^{2}\big)}\,,
\eeq
which is perfectly smooth at $T=0$ as long as $b\ne 0$
(see Fig.~\ref{fig:a-bounce-a-singular} for a comparison
with the singular solution).
The corresponding Kretschmann curvature scalar
$K \equiv R^{\mu\nu\rho\sigma}\,R_{\mu\nu\rho\sigma}$
and matter energy density $\rho$ are then finite at $T=0$,
provided $b \ne 0$,
\bsubeqs\label{eq:regularized-Friedmann-Ksol-rhosol}
\beqa
K(T)    &\propto& \big(b^{2}+T^{2}\big)^{-2}\,,
\\[2mm]
\rho(T) &\propto& \big(b^{2}+T^{2}\big)^{-1}\,.
\eeqa
\esubeqs
In terms of the auxiliary coordinate $\tau$
from \eqref{eq:mod-FLRW-tau-of-T-def},
the bounce solution reads
\beq
\label{eq:regularized-Friedmann-asol-tau}
a(\tau)\,\Big|_\text{mod.\;FLRW}^\text{(w=1,\;$\tau$-even\;sol.)} =
\sqrt[6]{\tau^{2}/\tau_{0}^{2}}\,,
\eeq
with $\tau_{0}^{2} \equiv b^{2}+T_{0}^{2}$.

We emphasize that the new input for this particular
nonsingular bouncing cosmology is
the metric \textit{Ansatz} \eqref{eq:mod-FLRW-ds2}.
The other two inputs are standard~\cite{Weinberg1972},
the Einstein equation and
the energy-momentum tensor of the matter
(here, matter with $w = 1$).
The resulting modified Friedmann equation
\eqref{eq:mod-Friedmann-equation-a},
together with the standard equations
\eqref{eq:mod-Friedmann-equation-b}
and \eqref{eq:mod-Friedmann-equation-c},
then gives the bounce-type scale
factor \eqref{eq:regularized-Friedmann-asol}.
In the next section, we calculate some
cosmological observables for this bounce-type FLRW universe.

%%\newpage%%tmp
\section{Cosmological observables}
\label{sec:Cosmological-observables}

\subsection{Null geodesics}
\label{subsec:Null-geodesics}

The background metric is given by \eqref{eq:mod-FLRW-ds2}.
Particles travel on straight lines in the coordinate system
$\{T,\,  x^{1},\, x^{2},\, x^{3}\}$.
So, we can consider geodesics of light that
start at $T = T_{1} < 0$ and end at $T = T_{0} > 0$, while
moving in the $x^1 \equiv X$ direction. Then, the reduced metric is
\beq\label{eq:mod.FLRW-reduced}
0=ds^{2} \,\Big|_{\text{mod. FLRW}} ^{(\text{light})}=
-\frac{T^{2}}{b^{2}+T^{2}} \,dT^{2}+a^{2}(T)  \,d X^{2} \,,
\eeq
where $c$ has been set to unity.
For matter with equation-of-state parameter $w = 1$,
the cosmic scale factor $a(T)$
is given by \eqref{eq:regularized-Friedmann-asol}.

\begin{figure}[t]
\vspace*{0mm}
\begin{center}
\includegraphics[width=0.5\textwidth]{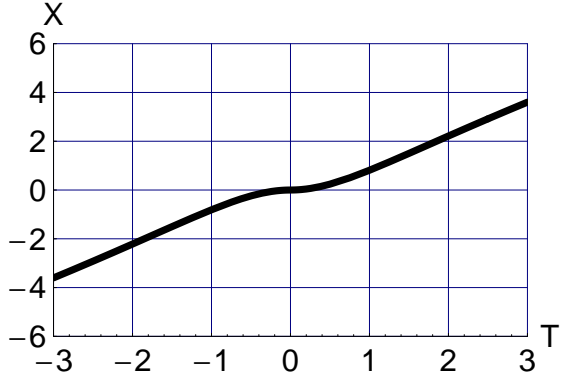}
%%{nonsingular-bouncing-cosmology-fig03-v380.eps}
%%{nonsingular-bouncing-cosmology-fig03-v3.eps}
%%{messengers-from-a-pre-BB-phase-fig03-v022.eps}
\end{center}
\vspace*{-5mm}
\caption{Null geodesic (\ref{eq:Xsol}) with $b=1$ and $T_{0}=4\,\sqrt{5}$.}
\label{fig:null-geodesic}
\end{figure}

With boundary condition $X(0) =0$,
we now have the following geodesic solution $X=X(T)$
from the reduced metric (\ref{eq:mod.FLRW-reduced})
and the cosmic scale factor \eqref{eq:regularized-Friedmann-asol}:
\beq
\label{eq:Xsol}
  X(T) =
 \begin{cases}
+\ddfracNEW{3}{2}\;\sqrt[6]{b^2+T_{0} ^2}\;
 \bigg[\sqrt[3]{T^2+b^2}-\sqrt[3]{b^2}\,\bigg]\,,
&\;\;\text{for}\;\;T > 0 \,,
 \\[4mm]
-\ddfracNEW{3}{2}\;\sqrt[6]{b^2+T_{0} ^2}\;
 \bigg[\sqrt[3]{T^2+b^2}-\sqrt[3]{b^2}\,\bigg]\,,
&\;\;\text{for}\;\;T \leq 0 \,.
 \end{cases}
\eeq
A plot of this null geodesic is given in Fig.~\ref{fig:null-geodesic}.

The conclusion is that particles, in particular photons and gravitons,
can travel from the prebounce phase to the postbounce
phase without hindrance whatsoever.

%%\newpage%%tmp
\subsection{Past particle horizon}
\label{subsec:Past-particle-horizon}

At cosmic time $T_{0}>0$, the past particle horizon
is infinite, as the universe extends back in time indefinitely.
Explicitly, the particle horizon at $T_{0}>0$ reads
\beq\label{eq:dhor-integrals}
d_\text{hor}(T_{0})= a(T_{0})\,\lim_{\tau_{1} \to -\infty}\;
\Bigg[\int_{\tau_{1}}^{-b} \frac{d\tau''}{a(\tau'')}
            + \int_{b}^{\tau(T_{0})} \frac{d\tau'}{a(\tau')} \Bigg]\,,
\eeq
where $\tau(T_{0})\equiv \tau_{0}$ is given by \eqref{eq:mod-FLRW-tau-of-T-def}
and  $a(\tau)$ by \eqref{eq:regularized-Friedmann-asol-tau}.
For positive and finite values of $b$ and $\tau_{0}$,
we get
\beqa
\label{eq:dhor-result}
d_\text{hor}(T_{0})&=& \frac{3}{2}\,a(T_{0})\,\lim_{\tau_{1} \to -\infty}\;
\bigg(\sqrt[3]{\tau_{1} ^2\,\tau_{0}}
- 2\,\sqrt[3]{b ^2\,\tau_{0}} + \tau_{0}
\bigg)
\nonumber\\[2mm]
&=& \frac{3}{2}\,a(T_{0})\,\lim_{\tau_{1} \to -\infty}\;
    \sqrt[3]{\tau_{1} ^2\,\tau_{0}}\,,
\eeqa
which goes to $+\infty$.
In other words, the past particle horizon at a finite positive cosmic
time $T_{0}$ diverges for this particular bounce-type universe.

With an infinite particle horizon, there may be
no horizon and flatness problems to worry about,
and no need for inflation~\cite{Guth1981}
(further references on inflationary models
can be found in Ref.~\cite{Mukhanov2005}).
A succinct comparison between nonsingular bouncing cosmology models
and big bang inflationary models appears in Ref.~\cite{IjjasSteinhardt2018}.

%%\newpage%%tmp
\subsection{Modified Hubble diagrams}
\label{subsec:Modified-Hubble-diagrams}

It is a straightforward exercise
to calculate the luminosity distance $d_{L}$
as a function of the redshift $z$, provided we distinguish two cases:
\begin{enumerate}
\item
the light is emitted by a comoving
source in the expanding phase of the universe ($T_{1} >0$);
\item
the light is emitted by a comoving source in the
contracting phase of the universe ($T_{1} \leq 0$).
\end{enumerate}
In both cases, the light is detected by a comoving observer in
the expanding phase at cosmic time $T_{0} >0$ with $T_{0} >T_{1}$.

Using the auxiliary time coordinate $\tau$
from \eqref{eq:mod-FLRW-tau-of-T-def}
with scale factor \eqref{eq:regularized-Friedmann-asol-tau}
and adapting the relevant formulae in
Secs.~14.4 and 14.6 of Ref.~\cite{Weinberg1972},
we obtain
\bsubeqs\label{eq:dLintegral-case1-case2}
\beqa\label{eq:dLintegral-case1}
\hspace*{-0mm}
d_{L}(\tau_{0},\,\tau_{1})\,\Big|^\text{(case\;1)}&=&
\frac{a^{2}(\tau_{0})}{a(\tau_{1})}\,\int_{\tau_{1}}^{\tau_{0}}\,
 \frac{d \tau'}{a(\tau')}\,,
\\[2mm]
\label{eq:dLintegral-case2}
\hspace*{-0mm}
d_{L}(\tau_{0},\,\tau_{1})\,\Big|^\text{(case\;2)}&\equiv&
d_{L}^\text{(pre)}(\tau_{1}) + d_{L}^\text{(post)}(\tau_{0})
\nonumber\\[1mm]
&=&
  \frac{a^{2}(-b)}{a(\tau_{1})}\,\int_{\tau_{1}}^{-b}\, \frac{d \tau''}{a(\tau'')}
+ \frac{a^{2}(\tau_{0})}{a(b)}\,\int_{b}^{\tau_{0}}\, \frac{d \tau'}{a(\tau')}\,,
\eeqa
\esubeqs
where light is emitted at cosmic time $\tau=\tau_{1}$
(with $\tau_{1}>b$ for case 1 and $\tau_{1}\leq -b$ for case 2)
and observed at $\tau=\tau_{0}>b>0$ with $\tau_{0}>\tau_{1}$.
Taking the positive function $a(\tau)$
from \eqref{eq:regularized-Friedmann-asol-tau}
and introducing the redshift,
\beq\label{eq:redshift}
z\equiv \sqrt{a^{2}(\tau_{0})/a^{2}(\tau_{1})}-1
  =      a(\tau_{0})/a(\tau_{1})-1\,,
\eeq
the integrals in \eqref{eq:dLintegral-case1-case2} give
\bsubeqs\label{eq:dL-case1-case2-zmax}
\beqa\label{eq:dL-case1}
\hspace*{-7mm}
d_L(z)\,\Big|^\text{(case\;1)}_{z \in [0,\,z_\text{max})}
&=&
3\,\tau_{0}\,\frac{1}{2}\,\left[ 1+z -\frac{1}{1+z}\right]\,,
\\[2mm]
\label{eq:dL-case2}
\hspace*{-7mm}
d_L(z)\,\Big|^\text{(case\;2)}_{z \in (-1,\,z_\text{max}]}
&=&
3\,\tau_{0}\,\frac{1}{2}\,
\left[ 1+z_\text{max} - \frac{1}{1+z_\text{max}}
+ \frac{1}{(1+z_\text{max})^2}
  \left( \frac{1}{1+z}-\frac{1+z}{(1+z_\text{max})^2}\right)\right] \,,
\nonumber\\[1mm]&&
\eeqa
with definition
\beqa
\label{eq:zmax}
z_\text{max} &\equiv& a(\tau_{0})/a(b)-1 =
\sqrt[3]{\tau_{0}/b}-1\,.
\eeqa
\esubeqs
The length scale $3\,\tau_{0}$ entering \eqref{eq:dL-case1}
and  \eqref{eq:dL-case2} is determined
by \eqref{eq:regularized-Friedmann-asol-tau},
\beq
\label{eq:2tau0-def}
3\,\tau_{0} =
\left[ \frac{1}{a(\tau)}\,\frac{d a(\tau)}{d \tau}
\right]_{\tau = \tau_{0}}^{-1}
\equiv \left[ H_{0}^{(\tau-\text{def.})}\right]^{-1}\,,
\eeq
where the Hubble constant $H_{0}^{(\tau-\text{def.})}$
differs from $H_{0}^{(T-\text{def.})}
\equiv [d a(T)/d T]/a(T)\,\big|_{T=T_{0}}$
by a factor close to unity, as long as $\tau_{0} \gg b$.

The corresponding expressions for
the angular diameter distance $d_{A}$ read
\bsubeqs\label{eq:dAintegral-case1-case2}
\beqa\label{eq:dAintegral-case1}
\hspace*{-3mm}
d_{A}(\tau_{0},\,\tau_{1})\,\Big|^\text{(case\;1)}&=&
\frac{a^{2}(\tau_{1})}{a^{2}(\tau_{0})}\,
d_{L}(z)\,\Big|^\text{(case\;1)}\,,
\\[2mm]
\label{eq:dAintegral-case2}
\hspace*{-3mm}
d_{A}(\tau_{0},\,\tau_{1})\,\Big|^\text{(case\;2)}
&=&
\frac{a^{2}(\tau_{1})}{a^{2}(-b)}\,d_{L}^\text{(pre)}(\tau_{1})
+
\frac{a^{2}(b)}{a^{2}(\tau_{0})}\,d_{L}^\text{(post)}(\tau_{0})\,.
\eeqa
\esubeqs
With the definitions in \eqref{eq:dLintegral-case1-case2}
and $a(\tau)$ from \eqref{eq:regularized-Friedmann-asol-tau},
the integrals give
\bsubeqs\label{eq:dA-case1-case2}
\beqa\label{eq:dA-case1}
\hspace*{-3mm}
d_A(z)\,\Big|^\text{(case\;1)}_{z \in [0,\,z_\text{max})}
&=&
3\,\tau_{0}\,\frac{1}{2}\,\frac{1}{(1+z)^2}\,
\left[ 1+z -\frac{1}{1+z}\right]\,,
\\[3mm]
\label{eq:dA-case2}
\hspace*{-3mm}
d_A(z)\,\Big|^\text{(case\;2)}_{z \in (-1,\,z_\text{max}]}
&=&
3\,\tau_{0}\,\frac{1}{2}\, \left[ \frac{1}{(1+z)^3}
-\frac{1}{(1+z_\text{max})^3}+\frac{z_\text{max}
+z\,(1+z_\text{max})}{(1+z_\text{max})^2(1+z)}\right]\,.
\eeqa
\esubeqs

\begin{figure}[t]
\vspace*{-4mm}
\begin{center}
\includegraphics[width=0.9\textwidth]{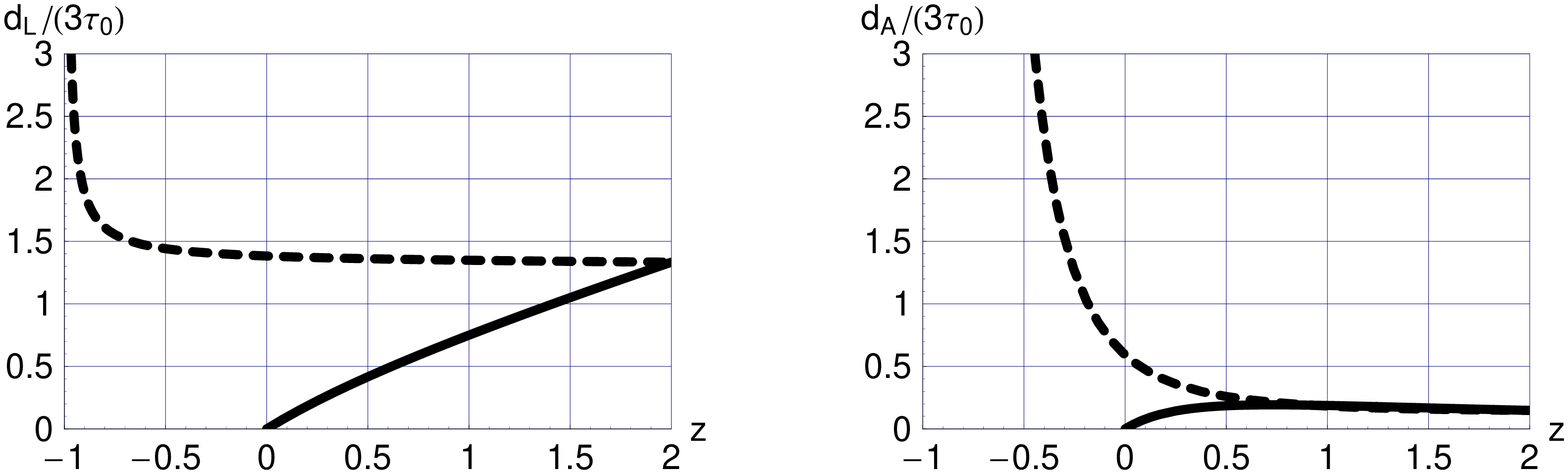}
%%{nonsingular-bouncing-cosmology-fig4-v396.eps}
%%{nonsingular-bouncing-cosmology-fig04-v4.eps}  %%NOT
%%{nonsingular-bouncing-cosmology-fig04-v380.eps}
%%{nonsingular-bouncing-cosmology-fig04-v3.eps}
%%{messengers-from-a-pre-BB-phase-fig04-v022.eps}
\end{center}
\vspace*{-6mm}
\caption{Modified Hubble diagrams for, on the left, the luminosity distance $d_{L}(z)$
from \eqref{eq:dL-case1-case2-zmax} and, on the right,
the angular diameter distance $d_{A}(z)$ from \eqref{eq:dA-case1-case2}.
The model parameters are $b/\tau_{0}=1/27$ and $z_\text{max}=2$.
With a comoving observer in the expanding phase,
the full curves correspond  %%FRK: v593
to case 1 (light emitted by a comoving source in the expanding phase
of the universe) and the dashed curves to case 2
(light emitted by a comoving source in the contracting phase).
}
\label{fig:modified-Hubble-dL-dA-diagram}
%\end{figure}
\vspace*{0mm}
%\begin{figure}[p]
\vspace*{-5mm}
\begin{center}
\includegraphics[width=0.9\textwidth]{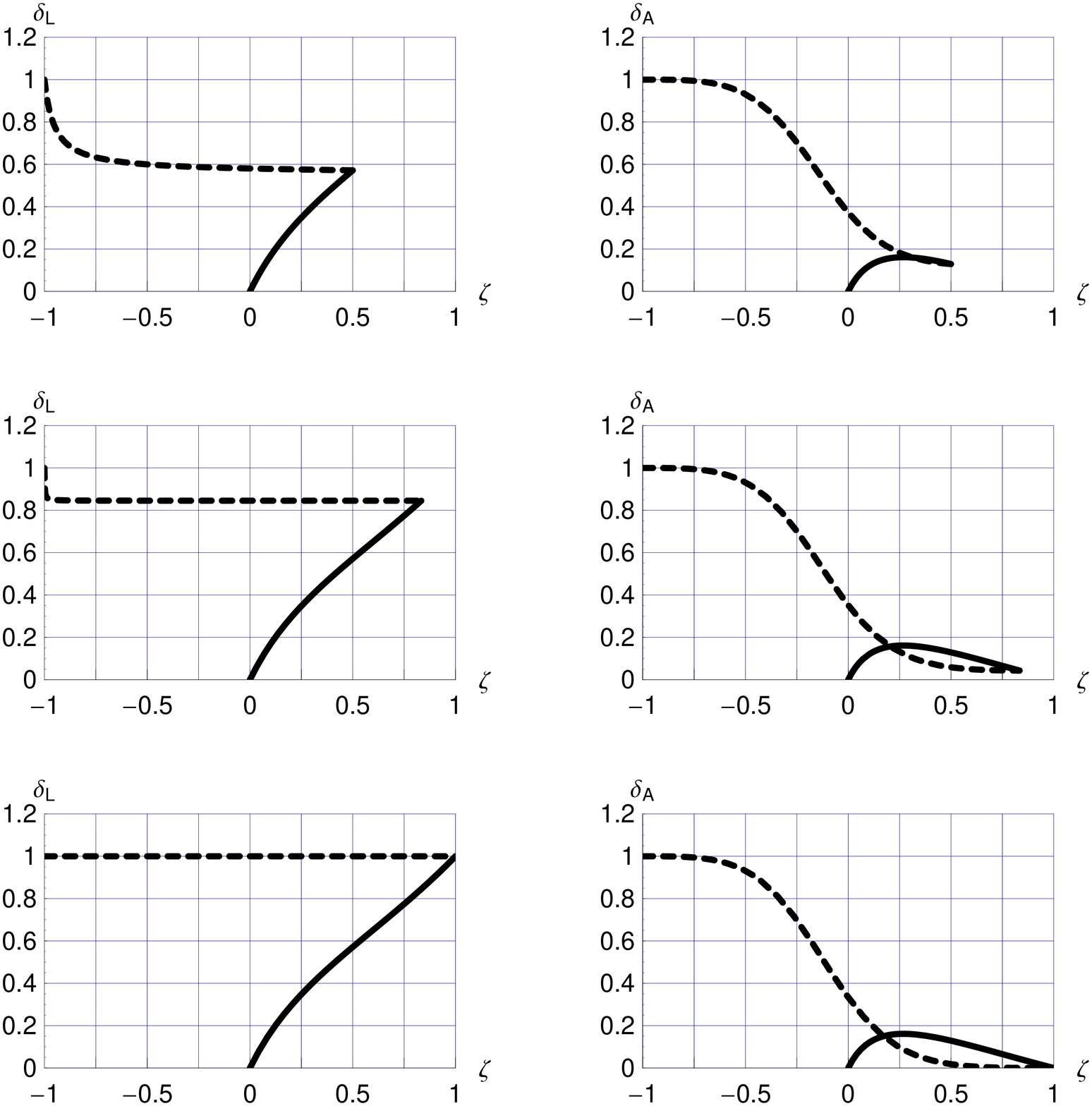}
\end{center}
\vspace*{-8mm}
\caption{Modified Hubble diagrams from \eqref{eq:dL-case1-case2-zmax}
and \eqref{eq:dA-case1-case2}, using the compactified redshift
variable $\zeta$ from \eqref{eq:zeta-def}
and the compactified distance variables
$\delta_L$ from \eqref{eq:deltaL-def}
and $\delta_A$ from \eqref{eq:deltaA-def}.
The top row has $z_\text{max}=2$,
the middle row $z_\text{max}=10$, and
the bottom row $z_\text{max}=10^{15}$.
The top-row curves correspond to
those of Fig.~\ref{fig:modified-Hubble-dL-dA-diagram}.
}
\label{fig:modified-Hubble-dL-dA-diagram-compactified}
\end{figure}

The modified Hubble diagram with the luminosity distance $d_{L}(z)$
is plotted in the left-panel of
Fig.~\ref{fig:modified-Hubble-dL-dA-diagram}
and the one with the angular diameter distance $d_{A}(z)$
in the right-panel.
The nonsmooth  behavior at $z=z_\text{max}$
in Fig.~\ref{fig:modified-Hubble-dL-dA-diagram}
is a direct manifestation of the spacetime defect and will
be discussed further in
Sec.~\ref{subsec:Cusps-in-the-Modified-Hubble-diagrams}.

The results in Fig.~\ref{fig:modified-Hubble-dL-dA-diagram} are
shown for a relatively small value of $z_\text{max}$, in order
to display the main characteristics of the modified Hubble diagrams.
For very large values of $z_\text{max}$ (as will appear in
Sec.~\ref{sec:Discussion}), it makes more sense to
use a compactified redshift coordinate and to compress
(or compactify) the distance axis.
Specifically, we can use the following compactified variables:
\bsubeqs\label{eq:zeta-deltaL-deltaA-def}
\beqa
\label{eq:zeta-def}
\zeta    &\equiv&  \frac{z}{z+2} \in (-1,\,1)\,,
\\[2mm]
\label{eq:deltaL-def}
\delta_L &\equiv&  \frac{d_L}{d_L + 3\,\tau_0} \in [0,\,1)\,,
\\[2mm]
\label{eq:deltaA-def}
\delta_A &\equiv&  \frac{d_A}{d_A + 3\,\tau_0} \in [0,\,1)\,.
\eeqa
\esubeqs
The resulting modified Hubble diagrams
are shown in Fig.~\ref{fig:modified-Hubble-dL-dA-diagram-compactified}
for three values of $z_\text{max}$.

After we completed our calculation of the
luminosity distance, we became aware
of Ref.~\cite{BarrauMartineauMoulin2017},
which discusses certain phenomenological aspects
of a nonsingular bouncing cosmology
but not the dynamics of the bounce.
The behavior of the $n=1/2$ curve in Fig.~1 of
Ref.~\cite{BarrauMartineauMoulin2017}
agrees with the more or less constant behavior of the
dashed curve in Fig.~4 of a previous version of this
article~\cite{KlinkhamerWang2019-preprint-v3},
which considered the $w=1/3$ case.

In the next two subsections, we will discuss these
modified Hubble diagrams in more detail.

%%\newpage%%tmp
\subsection{Cusps in the modified Hubble diagrams}
\label{subsec:Cusps-in-the-Modified-Hubble-diagrams}

The cusp behavior seen in Fig.~\ref{fig:modified-Hubble-dL-dA-diagram}
is of interest in that it shows that
the spacetime defect at $T=0$ (or $\tau=\pm b$)
is not just a coordinate artifact, as it leads to observable effects.
The discontinuity of the derivative $d^{\,\prime}_{L}(z)$
at $z_\text{max}$ from \eqref{eq:dL-case1} and \eqref{eq:dL-case2}
traces back to the nontrivial $\tau_{1}$ behavior
in \eqref{eq:dLintegral-case1-case2},
due to the sharp change in slope of
$a(\tau_{1})$ between $\tau_{1} \leq -b$ and $\tau_{1} \geq b$.
This change of slope is explained by two facts
(the overdot stands for differentiation with respect to $\tau$).
First, the
modified first-order Friedmann equation \eqref{eq:mod-Friedmann-equation-a}
in terms of the auxiliary coordinate $\tau$ implies that
the value of $(\dot{a}/a)^2$ at $\tau=\pm b$ is nonvanishing
if the value of $\rho$ is.
Second, the nonzero value of $\dot{a}/a$ can change sign between
$\tau=-b$ and $\tau=+b$, because the interval between these
two points ($\Delta\tau=2\,b$) is nonvanishing, as long as $b$ is nonzero.
Incidentally, we have also calculated $d_L(z)$
from \eqref{eq:dLintegral-case1-case2}
with an \emph{ad hoc} function $a(\tau)=1+(\tau^2/b^2-1)^2$
and find that the cusp in the
resulting modified Hubble diagram is absent.

The possible cusp behavior of the luminosity distance $d_{L}(z)$
has, to the best of our knowledge, not been obtained before in
other bouncing models. In Ref.~\cite{BarrauMartineauMoulin2017},
the authors did calculate the luminosity distances for different contracting 
phases but did not give a complete description,
from contraction to expansion.
Needless to say, a complete description of the luminosity distance is
far from trivial for most of the bouncing models in the literature,
as it depends on the details of the bouncing models
(especially the dynamics at the bounce moment).
As shown in Sec.~\ref{subsec:Modified-Hubble-diagrams},
our bouncing model not only gives a complete description
of the luminosity distance
(or the angular diameter distance) but also displays a nontrivial effect
such as the cusp behavior,
which may be regarded as a characteristic of our bouncing model.

%%\newpage%%tmp
\subsection{Gedankenexperiment}
\label{subsec:Gedankenexperiment}

Assume that modified Hubble diagrams for $d_L(z)$ and $d_A(z)$
have been established. Then, we may consider
a \textit{Gedankenexperiment} to determine
the numerical value of $b$ (presupposing the relevance
of the nonsingular bounce model of
Sec.~\ref{sec:Nonsingular bounce-with-a-constant-EOS}
to the real Universe).
The simplest possible \textit{Gedankenexperiment}
uses only the modified $d_L(z)$  Hubble diagram
and proceeds in three steps.

First, determine the numerical value of $z_\text{max}$
from the modified $d_L(z)$  Hubble diagram,
where the $z_\text{max}$ value corresponds to
the $z$-coordinate of the intersection point of the \mbox{case-1} curve
\eqref{eq:dL-case1} and the case-2 curve \eqref{eq:dL-case2};
compare with, respectively, the full and dashed curves in
the left-panel of Fig~\ref{fig:modified-Hubble-dL-dA-diagram}.
With the obtained $z_\text{max}$ value, calculate
\beq\label{eq:Gedankenexperiment-Xi-L}
\Xi_{L} \equiv  1+z_\text{max} \,.
\eeq
In this subsection, we use upper-case Greek letters
to denote experimental quantities.

Second, determine, in the modified $d_L(z)$ Hubble diagram,
the numerical value of the slope of the upper (case-2) curve
just below $z=z_\text{max}$,
\beq\label{eq:Gedankenexperiment-sL}
\Sigma_{L} \equiv
\frac{d}{d z}\,\left[  d_L(z)\,\Big|^\text{(case-2)}\right]_{z=z_\text{max}}
=\frac{-3}{1+z_\text{max}}\;b\,,
\eeq
where the last identification results from
\eqref{eq:dL-case2}. In principle,
it is also possible to obtain other experimental quantities
[for example, from the behavior of the curvature of the
case-2 $d_L(z)$ curve just above $z=-1$], but the choice
\eqref{eq:Gedankenexperiment-sL} suffices for the moment.

Third, the numerical value of $b$ then follows from the
previously determined values $\Xi_{L}$ and $\Sigma_{L}$
by calculating
\beq\label{eq:Gedankenexperiment-dnum}
b_\text{num} = - \frac{1}{3}\,\Sigma_{L} \; \Xi_{L}\,.
\eeq
With $z_\text{max}$ significantly larger than $1$,
the numerical value of $b$ in \eqref{eq:Gedankenexperiment-dnum}
results from multiplying a reduced value ($\Sigma_{L}$) by
a large number ($\Xi_{L}$). Note also  that the
$d_L(z)$ function obtained for the $w=1/3$
case~\cite{KlinkhamerWang2019-preprint-v3}
gives the same result as in \eqref{eq:Gedankenexperiment-dnum}
but with the fraction $1/3$ on the right-hand side replaced by $1/2$.

Needless to say, we neglect all practical difficulties
in this \textit{Gedankenexperiment}, if at all feasible.
A discussion of the cosmological context is given
in Sec.~\ref{sec:Discussion}.

%%\newpage%%tmp
\section{Discussion}
\label{sec:Discussion}

The construction of the spacetime manifold in Ref.~\cite{Klinkhamer2019}
is entirely classical. But it could very well be that the
classical length parameter $b$ appearing in the metric \eqref{eq:mod-FLRW-ds2}
has its origin in the (unknown) theory of ``quantum spacetime,''
with a fundamental length scale related to the
Planck length or not~\cite{Klinkhamer2007}.
It is, then, possible to imagine that this quantum theory removes
the classical times $\tau \in (-b,\, b)$
in Fig.~\ref{fig:sketch-tau-axis-with-T} and ties together
$\tau =-b$  and $\tau = b$, so that the resulting interval of the
emerging classical time coordinate $T=T(\tau)$ has no boundary.
In that case, there must be a classical ``pre-big-bang''
phase $T \leq 0$
and, in this article, we have studied some cosmological consequences
(one example being the cusps in the modified
Hubble diagrams of Fig.~\ref{fig:modified-Hubble-dL-dA-diagram},
as explained in Sec.~\ref{subsec:Cusps-in-the-Modified-Hubble-diagrams}).

Assuming the relevance of the nonsingular bounce model of
Sec.~\ref{sec:Nonsingular bounce-with-a-constant-EOS},
we have discussed, in Sec.~\ref{subsec:Gedankenexperiment},
a \textit{Gedankenexperiment} to determine
the numerical value of the length scale $b$ entering the
metric \eqref{eq:mod-FLRW-ds2}.
The required observations for this \textit{Gedankenexperiment}
would concern images originating \emph{before} the known epoch of the
``hot big bang'' (postbounce in our model universe),
which contains a hot plasma that would strongly scatter the light
of any assumed ``standard candles'' (or ``standard-size objects'')
in the prebounce phase.
But it may very well be that the required standard candles
emit gravitational waves instead of electromagnetic waves (light).
For example, it is possible to consider a
gravitational standard candle from a binary-black-hole
merger~\cite{LIGO-2016}
with definite masses (giving a recognizable ``chirp'');
see Ref.~\cite{BarrauMartineauMoulin2017} for further discussion.

Hence, the \textit{Gedankenexperiment} of
Sec.~\ref{subsec:Gedankenexperiment} relies on
gravitational standard candles.
The cosmological scenario for which this
\textit{Gedankenexperiment} may be relevant is as follows.
In the prebounce phase and just after the
bounce, the matter content of the universe can be
described by a homogeneous perfect fluid with
a constant equation-of-state parameter $w = 1$.
The particular value $w \geq 1$ agrees with the
absence of instabilities in the prebounce phase,
as discussed in the third and fourth paragraphs of
Sec~4 in Ref.~\cite{IjjasSteinhardt2018}, which also contains
further references.
After the bounce, the matter content of the universe
changes to that of a homogeneous perfect fluid with $w \sim 1/3$,
attributed to the presence of ultrarelativistic particles.
In the Appendix,  we present a model with a postbounce change of the equation-of-state parameter.

Even if the cosmological scenario as discussed holds true,
there are formidable hurdles to overcome before
the \textit{Gedankenexperiment} can be converted into a
realistic experiment. We only mention two.
The first major hurdle (perhaps the most important one)
concerns the actual presence and identification of
the required gravitational standard candles
(or gravitational standard-size objects),
present before and after the bounce.
The second major hurdle concerns the measurement of $b$
by use of \eqref{eq:Gedankenexperiment-dnum}.
Using the known postbounce age of the universe
($c\,\tau_{0} \approx 10^{10}\,\text{lyr}\approx 10^{26}\,\text{m}$)
and taking the maximum allowed value for $b$ from
\eqref{eq:upper-bound-on-b},
we get the following estimate from \eqref{eq:zmax}
adapted to the postbounce expansion $a(\tau) \propto \tau^{1/2}$
for the model of the Appendix:
\beqa
\label{eq:zmax-numerical}
1+ z_\text{max} \sim
10^{15}\;\left(\frac{c\,\tau_{0}}{10^{26}\,\text{m}}\right)^{1/2}
\;\left(\frac{10^{-3}\,\text{m}}{b}\right)^{1/2}\,.
\eeqa
This large number $10^{15}$
(or an even larger number if, for example,
$b$ is very much below the millimeter scale) makes the determination
of $b$ difficult, as mentioned in the sentence below
\eqref{eq:Gedankenexperiment-dnum}.
The slope entering \eqref{eq:Gedankenexperiment-sL},
in particular, is suppressed by, at least, a factor $10^{-15}$,
making it hard to measure.

The experiment as outlined above will stay,
for a long time to come,
a \textit{Gedankenexperiment} and the discussion
will remain entirely academic.
Still, the general idea appears to be valid
and may perhaps be adapted to other circumstances.

Up till now, we have been talking primarily about direct images
of prebounce structures (e.g., hypothetical binary-black-hole mergers
emitting gravitational waves).
But, as mentioned in Fig.~4 of Ref.~\cite{IjjasSteinhardt2018},
the currently observed ``superhorizon''
patterns in the cosmic microwave background
may also be due to a prebounce phase,
assuming that there has been such a phase.
Hence, the crucial question is whether or not a cosmic bounce
has occurred and, if so, what physics is responsible.
The intriguing result from
general relativity, extended to allow for degenerate metrics,
is that a particular ``regularization'' of the
standard big bang singularity  indeed suggests the
occurrence of a cosmic bounce.

\vspace*{-2mm}
\begin{acknowledgments}
\vspace*{-2mm}
We thank the referee for useful comments.
The work of Z.L.W. is supported by the China Scholarship Council.
\end{acknowledgments}

\vspace*{-2mm}
\section*{Note Added}
\vspace*{-2mm}

The present article is a follow-up paper
of Ref.~\cite{Klinkhamer2019}.
There are now two more follow-up papers.
The first of these papers~\cite{KlinkhamerWang2019-lhep}
provides a scalar-field  model
for the type of time-asymmetric nonsingular bounce
constructed in the Appendix.
The second of these papers~\cite{Klinkhamer2019-revisited}
gives a detailed analysis of the dynamics near the bounce.

%%\newpage%%tmp
\begin{appendix}
\section{Nonsingular bounce with a variable equation of state}
\label{app:Nonsingular bounce-with-a-variable-EOS}

In this Appendix,  we give some results for a
modified spatially flat FLRW universe with a
time-dependent equation-of-state parameter.
In fact, we take our cue from the general discussion of a
particular classical nonsingular bouncing cosmology
in Ref.~\cite{IjjasSteinhardt2018}.
With the notation $\epsilon \equiv (3/2)(1+P/\rho)\equiv (3/2)(1+W)$,
Sec.~4 of that paper states:
``According to the bouncing scenario, at some point during
or shortly after the bounce,
the kinetic energy stored in scalar fields
is converted to the matter and radiation we observe,
with $\epsilon \leq 2$. The irreversible reheating process accounts
for the asymmetry in $\epsilon$ about the bounce point.''
The main characteristics of that nonsingular bouncing cosmology
are summarized in Fig.~3 of Ref.~\cite{IjjasSteinhardt2018}
and the goal of the present Appendix  is to present
a ``fully-computable bounce model,'' as mentioned
in Sec.~6 of Ref.~\cite{IjjasSteinhardt2018}.

\begin{figure}[t!]
\vspace*{-5mm}
\begin{center}
\includegraphics[width=0.9\textwidth]{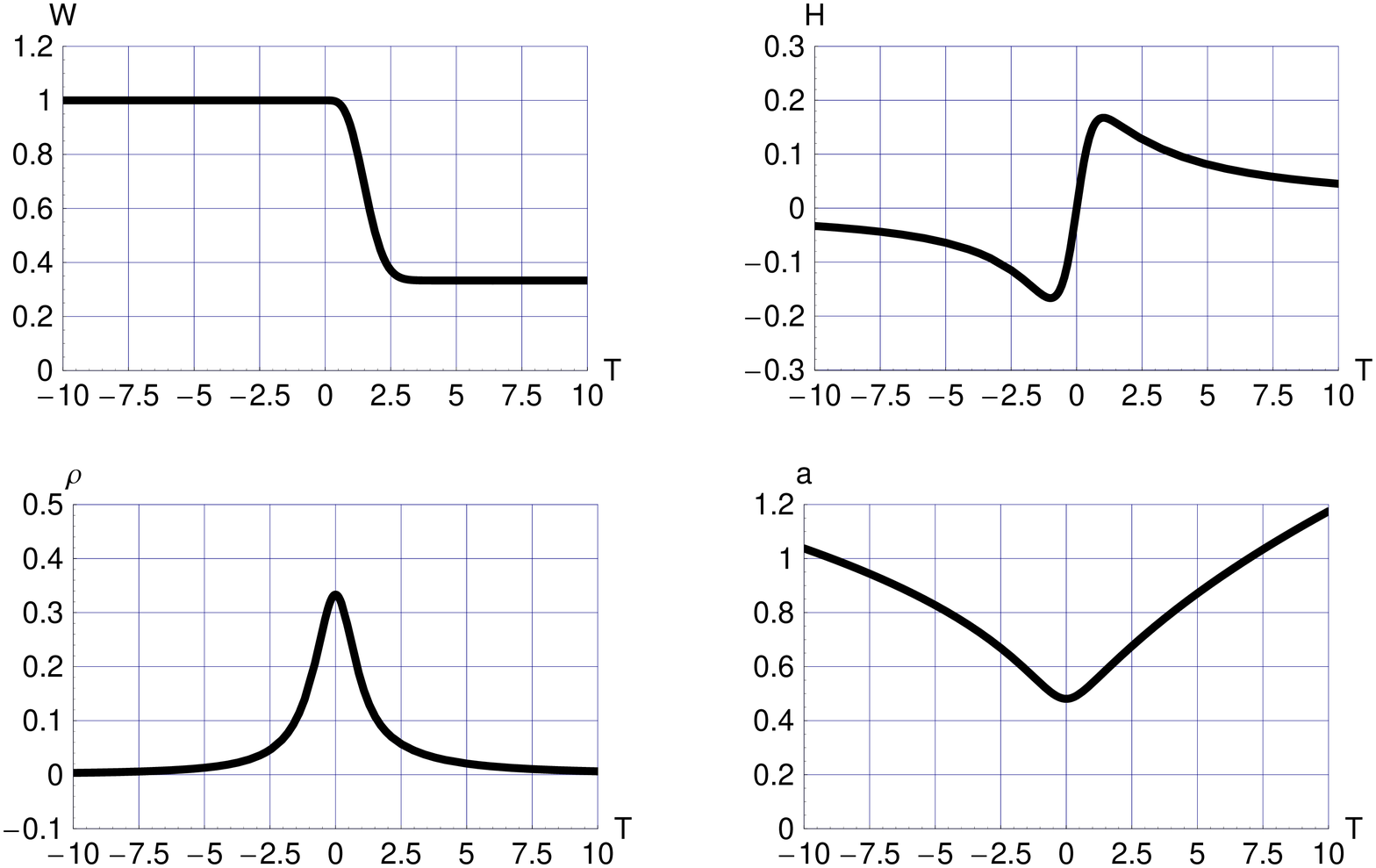}
%%{nonsingular-bouncing-cosmology-fig5-v4.eps}
%%{messengers-from-a-pre-BB-phase-fig06-v082.eps}
\end{center}
\vspace*{-5mm}
\caption{Bounce-type universe from the modified Friedmann
equation \eqref{eq:bounce-mod-Friedmann-equation-a}
with a postbounce change of the equation of
state \eqref{eq:bounce-mod-Friedmann-equation-c}.
Top-left panel: equation-of-state parameter $W(T)$
from \eqref{eq:bounce-mod-Friedmann-equation-d}.
Top-right panel: Hubble parameter $H(T)$ from
\eqref{eq:HsolT} and \eqref{eq:hbarsol}.
Bottom-left panel: energy density $\rho(T)$
from \eqref{eq:rhosolT}  and \eqref{eq:hbarsol}.
Bottom-right panel: numerical solution for the
cosmic scale factor $a(T)$ from
the ordinary differential equation \eqref{eq:HsolT}
with boundary condition $a(-T_{0})=1$.
The time-asymmetric behavior of $a(T)$ in the bottom-right panel
is manifest [having, for example, $a(10)\ne a(-10)$] and
differs from the symmetric behavior in Fig.~\ref{fig:a-bounce-a-singular}.
The model parameters
are $\big\{b,\,\tau_{0},\,T_{0}\big\}=\big\{1,\,9,\,4\,\sqrt{5}\, \big\}$
in reduced-Planckian units with $8\pi G_N=c=\hbar=1$.
}
\label{fig:bounce-universe}
\vspace*{-0mm}
%\end{figure}
\vspace*{5mm}
%\begin{figure}[h!]
\vspace*{-0mm}
\begin{center}
\includegraphics[width=0.9\textwidth]{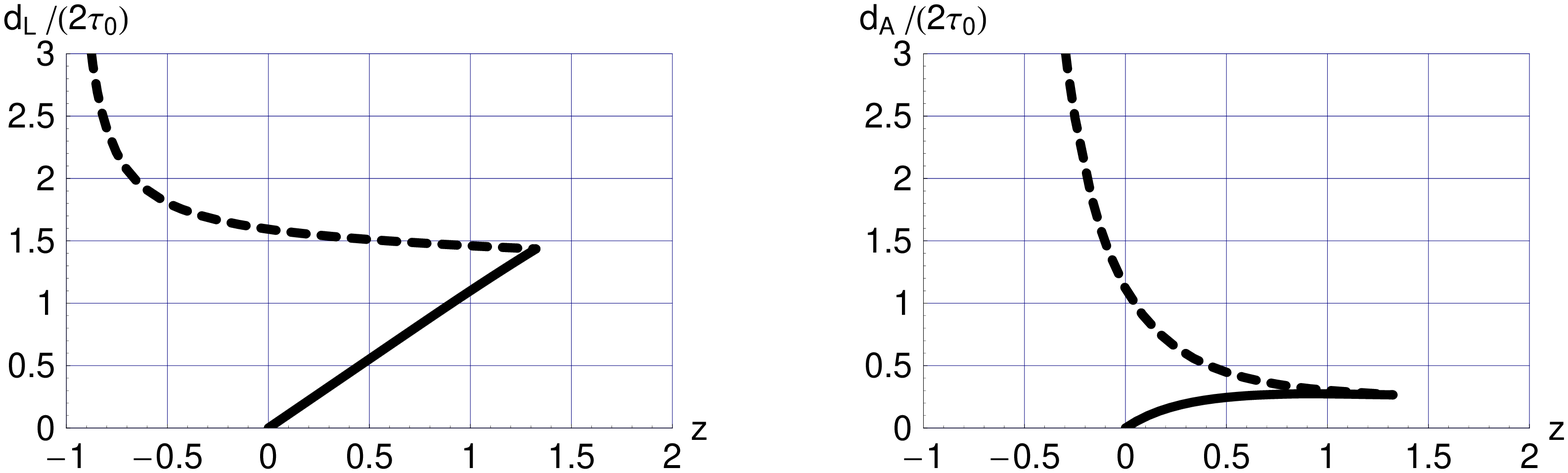}
%%{nonsingular-bounce-cosmology-fig07-v180.eps}
\end{center}
\vspace*{-5mm}
\caption{Modified Hubble diagrams based on numerical results for the
luminosity distance $d_{L}$ from \eqref{eq:dLintegral-case1-case2}
and the angular diameter distance $d_{A}$ from \eqref{eq:dAintegral-case1-case2}
in the bounce-type universe of Fig.~\ref{fig:bounce-universe}.
For the model parameters chosen, the maximum redshift
is given by $z_\text{max} \equiv a(T_{0})/a(0)-1 \approx 1.32425$.}
\label{fig:bounce-universe-dL-dA}
\vspace*{1000mm}
\end{figure}

With reduced-Planckian units ($8\pi G_N=c=\hbar=1$),
the modified spatially flat Friedmann
equation, the energy-conservation equation, and
the assumed equation of state are given by%
\bsubeqs\label{eq:bounce-mod-Friedmann-equations-abcd}
\beqa\label{eq:bounce-mod-Friedmann-equation-a}
\hspace*{-0mm}
\left(1+ \frac{b^{2}}{T^{2}}\right)
\left( \frac{1}{a(T)}\,\frac{d a(T)}{d T} \right)^{2}
&=& \frac{1}{3}\;\rho(T)\,,
%\\[2mm]
\eeqa\beqa
\label{eq:bounce-mod-Friedmann-equation-b}
\hspace*{-0mm}
\frac{d}{d a} \bigg[ a^3\,\rho(a)\bigg]+ 3\, a^{2}\,P(a)&=&0\,,
%\\[2mm]
\eeqa\beqa
\label{eq:bounce-mod-Friedmann-equation-c}
\hspace*{-0mm}
P(T) &=& W(T)\,\rho(T)\,,
\\[2mm]
\label{eq:bounce-mod-Friedmann-equation-d}
\hspace*{-0mm}
W(T) &=&
\begin{cases}
 \ddfracNEW{1}{3}+\ddfracNEW{2}{3}\,
  \exp\left[-\left(\sqrt{1+T^{2}/b^{2}}-1\right)^{2}\,\right]\,,
 &  \;\text{for}\; T > 0\,,
 \\[2mm]
 1\,,   &   \;\text{for}\; T \leq 0\,,
\end{cases}
\eeqa
\esubeqs
where \eqref{eq:bounce-mod-Friedmann-equation-c}
and \eqref{eq:bounce-mod-Friedmann-equation-d}
provide an explicit realization of the
required equation-of-state behavior of
the nonsingular bouncing cosmology as
displayed in Fig.~3 of Ref.~\cite{IjjasSteinhardt2018}.
The particular function $W(T)$ from \eqref{eq:bounce-mod-Friedmann-equation-d}
is shown, for model parameter $b=1$,
in the top-left panel of Fig.~\ref{fig:bounce-universe}.

By reverting temporarily to the auxiliary coordinate $\tau$ from
\eqref{eq:mod-FLRW-tau-of-T-def}
and by focusing on the Hubble parameter
$h(\tau)\equiv a^{-1}(\tau)\,[d a(\tau)/d \tau]$
it is possible to get an analytic result:%
\bsubeqs\label{eq:HsolT-rhosolT-hbarsol}
\beqa
\label{eq:HsolT}
\hspace*{-5mm}
H(T)
&\equiv&  \left[ \frac{1}{a(T)}\,\frac{d a(T)}{d T}\right]
=          \sqrt{\frac{T^{2}}{b^{2}+T^{2}}}\;\overline{h}(T) \,,
\\[2mm]
\label{eq:rhosolT}
\hspace*{-5mm}
\rho(T)&=& 3\,\overline{h}^{\,2}(T)\,,
\\[2mm]
\label{eq:hbarsol}
\hspace*{-5mm}
\overline{h}(T)&=&
\begin{cases}
 \Big(b+2\,\sqrt{b^{2}+T^{2}}
             +\ddfracNEW{1}{2}\,b\,\sqrt{\pi}\;
\text{erf}\left[ \sqrt{1+T^{2}/b^{2}}-1 \right]\Big)^{-1}\,,
 &  \;\;\text{for}\;\; T > 0\,,
\\[2mm]
 \Big(-3\,\sqrt{b^{2}+T^{2}}\Big)^{-1}\,,
 &  \;\;\text{for}\;\; T \leq 0\,,
\end{cases}
\eeqa
\esubeqs
in terms of the error function
\beq
\label{eq:erf-def}
\text{erf}(z)
\equiv
\frac{2}{\sqrt{\pi}}\,\int_{0}^{z} d t \, \exp\big(-t^{2}\big)\,.
\eeq
From \eqref{eq:HsolT} and \eqref{eq:hbarsol}, we have
$H(T) \sim (1/3)\,T^{-1} < 0$ for $T \ll -b$ and
$H(T) \sim (1/2)\,T^{-1} > 0$ for $T \gg b$.

It does not appear possible to get $a(T)$ in an explicit analytic form,
but the ordinary differential equation from \eqref {eq:HsolT}
can be solved numerically for $a(T)$.
Figure~\ref{fig:bounce-universe} shows the cosmological
functions for a particular choice of model parameters,
where the bottom-right panel displays the time asymmetry
of the cosmic scale factor, $a(T)\ne a(-T)$ for $T\ne 0$.
The corresponding luminosity distance $d_{L}$
and angular diameter distance $d_{A}$ (Fig.~\ref{fig:bounce-universe-dL-dA})
are found to be qualitatively the same as those
from Sec.~\ref{sec:Cosmological-observables}
(Fig.~\ref{fig:modified-Hubble-dL-dA-diagram}).

From \eqref{eq:rhosolT} and \eqref{eq:hbarsol},
we find that the maximum value of the energy density
(which occurs at $T = 0$) remains finite,
provided the defect length scale $b$ is nonzero
\beq
\label{eq:rho-maximum}
\rho(0) = \ddfracNEW{1}{3}\, E^{2}_\text{planck}\,b^{-2}\,,
\eeq
in terms of the reduced Planck energy,
\beq\label{eq:Eplanck-def}
E_\text{planck} \equiv \sqrt{\hbar\, c^5/(8\pi G_N)}
\approx 2.44 \times 10^{18}\,\text{GeV}\,.
\eeq
Demanding
\bsubeqs
\label{eq:lower-bound-on-rho-upper-bound-on-b}
\beqa
\label{eq:lower-bound-on-rho}
\rho(0) \gtrsim (\text{TeV})^4\,,
\eeqa
in order to reproduce, in the postbounce phase,
the hot-big-bang model with temperatures $\mathcal{T} \lesssim \text{TeV}$,
we have the following qualitative upper bound on
the defect length scale $b$ from \eqref{eq:rho-maximum}:
\beqa
\label{eq:upper-bound-on-b}
b \lesssim \left(\frac{E_\text{planck}}{\text{TeV}}\right)\,\hbar c/\text{TeV}
  \approx 10^{15}\,\hbar c/\text{TeV}\approx 10^{-3}\,\text{m}\,.
\eeqa
\esubeqs
The millimeter scale has, of course,  appeared before in
higher-dimensional TeV gravity~\cite{ArkaniHamedDimopoulosDvali1998},
essentially tracing back to the Einstein equation
[which, here, gives rise to \eqref{eq:rho-maximum}].

For the record, we can mention that we also
have a qualitative lower bound on the defect length scale $b$.
Demanding
\bsubeqs
\label{eq:upper-bound-on-rho-lower-bound-on-b}
\beqa
\label{eq:upper-bound-on-rho}
\rho(0) \lesssim (E_\text{planck})^4\,,
\eeqa
in order that the classical Einstein theory be applicable,
we have the following qualitative  lower bound on $b$
from \eqref{eq:rho-maximum}:
\beqa
\label{eq:lower-bound-on-b}
b \gtrsim  \hbar c/E_\text{planck} \equiv l_\text{planck}
  \approx 8.10 \times 10^{-35}\,\text{m}\,.
\eeqa
\esubeqs
Such a minimal length scale is also expected, on general grounds,
for the emerging classical spacetime~\cite{AshtekarSingh2011}.

\end{appendix}


\newpage
\begin{thebibliography}{99}

\bibitem{GasperiniVeneziano2002}
M.~Gasperini and G.~Veneziano,
``The pre-big-bang scenario in string cosmology,''
Phys. Rep.  {\bf 373}, 1 (2003),
%%doi:10.1016/S0370-1573(02)00389-7
arXiv:hep-th/0207130.
%%CITATION = doi:10.1016/S0370-1573(02)00389-7;%%

\bibitem{AshtekarSingh2011}
A.~Ashtekar and P.~Singh,
``Loop quantum cosmology: A status report,''
Class. Quant. Grav.  {\bf 28}, 213001 (2011),
%%doi:10.1088/0264-9381/28/21/213001
arXiv:1108.0893.  %%[gr-qc]].
%%CITATION = doi:10.1088/0264-9381/28/21/213001;%%


\bibitem{IjjasSteinhardt2018}
A.~Ijjas and P.J.~Steinhardt,
``Bouncing cosmology made simple,''
Class. Quant. Grav.  {\bf 35},  135004 (2018),
%%doi:10.1088/1361-6382/aac482
arXiv:1803.01961. %% [astro-ph.CO]].
%%CITATION = doi:10.1088/1361-6382/aac482;%%

\bibitem{Klinkhamer2019}
F.R.~Klinkhamer,
``Regularized big bang singularity,''
Phys. Rev. D {\bf 100}, 023536 (2019),
arXiv:1903.10450. %% [gr-qc].
%%CITATION = ARXIV:1903.10450;%%

\bibitem{Klinkhamer2014-mpla}
F.R.~Klinkhamer,
``A new type of nonsingular black-hole solution in general relativity,''
Mod. Phys. Lett. A {\bf 29}, 1430018 (2014),
arXiv:1309.7011.  %% [gr-qc]].
%%CITATION = doi:10.1142/S0217732314300183;%%


\bibitem{KlinkhamerSorba2014}
F.R.~Klinkhamer and F.~Sorba,
``Comparison of spacetime defects which are homeomorphic but not diffeomorphic,''
J. Math. Phys. (N.Y.)  {\bf 55}, 112503 (2014),
arXiv:1404.2901.  %%[hep-th]].
%%CITATION = doi:10.1063/1.4900883;%%


\bibitem{Guenther2017}
M. Guenther,
``Skyrmion spacetime defect, degenerate metric,
and negative gravitational mass,''
Master Thesis, KIT, September 2017;
available from
\verb"https://www.itp.kit.edu/en/"
\verb"publications/diploma"


\bibitem{Klinkhamer2007}
F.R.~Klinkhamer,
``Fundamental length scale of quantum spacetime foam,''
JETP Lett.  {\bf 86}, 73 (2007),
%%doi:10.1134/S0021364007140019
arXiv:gr-qc/0703009.
%%CITATION = doi:10.1134/S0021364007140019;%%

\bibitem{Weinberg1972}
S.~Weinberg,
\emph{Gravitation and Cosmology : Principles and
Applications of the General Theory of Relativity}
(John Wiley and Sons, New York, 1972).
%%CITATION = INSPIRE-1410180;%%

\bibitem{Guth1981}
A.H.~Guth,
``The inflationary universe: A possible solution to the horizon
  and flatness problems,''
Phys. Rev. D {\bf 23}, 347 (1981).

\bibitem{Mukhanov2005}
V.~Mukhanov,
\emph{Physical Foundations of Cosmology}
%\href{http://www.slac.stanford.edu/spires/find/hep/www?irn=6927394}{SPIRES entry}
(Cambridge University Press, Cambridge, England, 2005).

\bibitem{BarrauMartineauMoulin2017}
A.~Barrau, K.~Martineau, and F.~Moulin,
``Seeing through the cosmological bounce: Footprints of the contracting
phase and luminosity distance in bouncing models,''
  Phys. Rev. D {\bf 96}, 123520 (2017),
%%doi:10.1103/PhysRevD.96.123520
arXiv:1711.05301. %% [gr-qc]].
%%CITATION = doi:10.1103/PhysRevD.96.123520;%%


\bibitem{LIGO-2016}
B.P.~Abbott {\it et al.} (LIGO Scientific and Virgo Collaborations),
``Observation of gravitational waves from a binary black hole merger,''
Phys. Rev. Lett.  {\bf 116}, 061102 (2016),
arXiv:1602.03837.  %% [gr-qc]].
%%CITATION = doi:10.1103/PhysRevLett.116.061102;%%

\bibitem{KlinkhamerWang2019-preprint-v3}
F.R.~Klinkhamer and Z.L.~Wang,
``Nonsingular bouncing cosmology from general relativity,''
arXiv:1904.09961v3. %% [gr-qc].
%%CITATION = ARXIV:1904.09961;%%

\bibitem{KlinkhamerWang2019-lhep}
F.R.~Klinkhamer and Z.L.~Wang,
``Asymmetric nonsingular bounce from a dynamic scalar field,''
Lett. High Energy Phys. {\bf 3}, 9 (2019),
arXiv:1906.04708.  %% [gr-qc].
%%CITATION = ARXIV:1906.04708;%%

\bibitem{Klinkhamer2019-revisited}
F.R.~Klinkhamer,
``Nonsingular bounce revisited,''
arXiv:1907.06547.  %% [gr-qc].
%%CITATION = ARXIV:1907.06547;%%


\bibitem{ArkaniHamedDimopoulosDvali1998}
N.~Arkani-Hamed, S.~Dimopoulos, and G.R.~Dvali,
``The hierarchy problem and new dimensions at a millimeter,''
Phys. Lett. B {\bf 429}, 263 (1998),
arXiv:hep-ph/9803315.
%%CITATION = doi:10.1016/S0370-2693(98)00466-3;%%


\end{thebibliography}
\end{document}